\documentclass[10pt,journal,compsoc]{IEEEtran}

\ifCLASSOPTIONcompsoc
  \usepackage[nocompress]{cite}
\else
  \usepackage{cite}
\fi

\usepackage{mathtools}

\usepackage{amsmath,amssymb,amsfonts}
\usepackage{algorithmic}
\usepackage{graphicx}
\usepackage{float}
\usepackage{subcaption}
\usepackage{multirow}
\usepackage{tabularx}

\begin{document}

\title{Optimal Multi-Level Interval-based Checkpointing for Exascale Stream Processing Systems\\
}

\author{Sachini Jayasekara,
        Aaron Harwood,
        and~Shanika Karunasekera% <-this % stops a space
\IEEEcompsocitemizethanks{\IEEEcompsocthanksitem S. Jayasekara, A. Harwood, and S. Karunasekera are with the School of Computing and Information Systems, The University of Melbourne, Melbourne, VIC 3010, Australia.\protect\\
E-mail: wjayasekara@student.unimelb.edu.au,aharwood@unimelb.edu.au, and karus@unimelb.edu.au}% <-this % stops an unwanted space
}

\IEEEtitleabstractindextext{%
\begin{abstract}
State-of-the-art stream processing platforms make use of checkpointing to support fault tolerance, where a ``checkpoint tuple" flows through the topology to all operators, indicating a checkpoint and triggering a checkpoint operation. The checkpoint will enable recovering from any kind of failure, be it as localized as a process fault or as wide spread as power supply loss to an entire rack of machines. As we move towards Exascale computing, it is becoming clear that this kind of ``single-level" checkpointing is too inefficient to scale. Some HPC researchers are now investigating multi-level checkpointing, where checkpoint operations at each level are tailored to specific kinds of failure to address the inefficiencies of single-level checkpointing. Multi-level checkpointing has been shown in practice to be superior, giving greater efficiency in operation over single-level checkpointing. However, to date there is no theoretical basis that provides optimal parameter settings for an interval-based coordinated multi-level checkpointing approach. This paper presents a theoretical framework for determining optimal parameter settings in an interval-based multi-level periodic checkpointing system, that is applicable to stream processing. Our approach is stochastic, where at a given checkpoint interval, a level is selected with some probability for checkpointing. We derive the optimal checkpoint interval and associated optimal checkpoint probabilities for a multi-level checkpointing system, that considers failure rates, checkpoint costs, restart costs and possible failure during restarting, at every level. We confirm our results with stochastic simulation and practical experimentation.
\end{abstract}

%% Note that keywords are not normally used for peerreview papers.
\begin{IEEEkeywords}
Fault tolerance, multi-level checkpoint, stream processing, optimization.
\end{IEEEkeywords}
}

\maketitle

\IEEEraisesectionheading{\section{Introduction}\label{sec:introduction}}

\begin{figure}[t]
\centering
\includegraphics[width=3in]{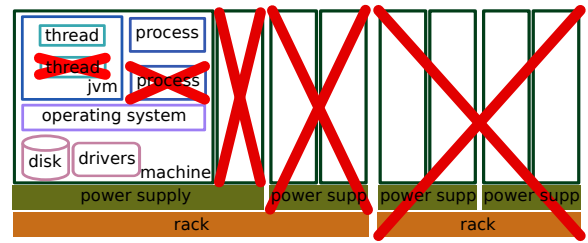}
\caption{Different types of failure levels could include, from right to left: rack failure, power supply failure, machine failure (including e.g. disk drive failure), process failure, thread failure. Failures at the rack and power supply level are expected to happen (very much) less frequently compared to failures at the process and thread levels.}
\label{fig:FL}
\end{figure}

Entering the Exascale computing era presents a tremendous challenge for stream processing platforms with respect to fault tolerance. In Exascale computing systems, failure rates are observed in the range of one failure every few minutes~\cite{5645453,1303239}. While this may seem implausible, consider that these rates include failures at every level of the system, from threads and processes to individual machines, groups of machines attached to a single power supply, and even entire racks, depicted in Figure~\ref{fig:FL}. For example, failures caused by unhandled exceptions, out of memory and buffer overflow issues can affect several, but not all, processes of a machine while hardware failures affecting several machines can have more severe impact. In some high performance computing (HPC) systems, sets of machines are combined to form a mid-plane and a rack is formed by a set of mid-planes. Failures affecting this hardware can have varying forms of impacts. For instance, a mid-plane switch failure can affect all of the machines in the mid-plane while the rack's power supply failure can affect all of the machines in the rack~\cite{Snir:2014:AFE:2747699.2747701,Herault:2015:FTH:2811302}. When a component fails, we assume that all of the state maintained by that component is lost. While the failure rate of any single component can be quite low, the aggregate failure rate readily becomes overwhelming.

Checkpointing is a strategy to overcome failure. In a single-level checkpointing approach, there is only one kind of checkpoint used over the entire system. Stream processing applications often referred to as topologies, process tuple streams using a set of operators and employ global checkpointing. Global checkpointing is achieved by flowing a ``checkpoint tuple" through the topology to all operators, indicating a checkpoint and triggering a checkpoint operation. When the tuple exits the topology then the checkpoint is complete. Failure of any kind involves restarting the topology from the latest checkpoint and replaying lost messages at the inputs to the topology. The only parameter to optimize is the \emph{checkpoint interval} and there is significant existing theoretical and practical results in this regard~\cite{CastroFernandez:2013:ISO:2463676.2465282,Daly:2006:HOE:1134241.1708449}. State-of-the-art stream processing systems such as \emph{Apache Flink} and \emph{Storm}~\cite{Carbone2015ApacheFS,Toshniwal:2014:STO:2588555.2595641} use the checkpoint/restart model~\cite{Carbone:2017:SMA:3137765.3137777,Low:2012:DGF:2212351.2212354} to support fault tolerance for stateful streaming applications~\cite{CastroFernandez:2013:ISO:2463676.2465282,5958214}. Replication~\cite{7164919,6267921} is another approach used to support fault tolerance. However, replication requires a large number of additional resources and becomes less effective as the state size grows~\cite{7164919}. 

Some HPC researchers have recently provided experimental results that show how checkpointing can be improved by considering different ``levels" of failure, especially in Exascale systems that are vulnerable to several different failure types~\cite{5645453,6569798,7795220}. Each level has its own failure rate, checkpoint cost and restart cost. Multi-level checkpoints can be performed in several ways. One approach is, each level independently performs checkpoints at its own unique checkpoint interval. This method has several issues such as overlapping checkpoints and is considered challenging to be applied in real-world systems~\cite{8425492,7795220}. Another approach is to define a repeating pattern of checkpoints over the levels~\cite{7795220}, e.g. for 2 levels: $1,1,2,1,1,2,1,1,2,\dotsc$ etc, which is modeled for bounded workloads. In this approach only the pattern is repeated periodically. However, inside the pattern, different types of checkpoints can happen one after another with no computations in-between leading to aperiodic checkpoint intervals. Unlike the pattern-based model, the model we are proposing is for systems that use periodic checkpoints, where all the checkpoints are performed periodically based on a unique checkpoint interval. Our model is similar to the single-level periodic checkpointing approach used in existing system with the exception of having different levels of checkpoints instead of one.

In this paper we talk about \emph{low level} checkpointing being a checkpoint that can recover from thread and process failure, while \emph{high level} checkpointing is one which can recover from more severe failure such as power supply and rack failure. A high level checkpoint can be used to recover from any of the failures at its own level or below, but this is not the case in the opposite direction. E.g. a thread or process can checkpoint to an in-memory checkpoint manager: checkpointing and recovery can be quite cheap but failure of the machine will need a higher level checkpoint to recover. Higher level checkpointing is more costly both in the time to save the checkpoint data and in the recovery, but higher level failures happen less often than lower level failures. Therefore the natural question in a multi-level checkpoint scheme is how often should we checkpoint at each level? However, to date there is no theoretical basis that provides optimal parameter settings for coordinated interval-based multi-level checkpointing approach with no overlapping checkpoints~\cite{8425492}. Existing approaches for multi-level use pattern-based checkpointing focusing on bounded workloads and the interval-based approach suggested only for two-level checkpointing~\cite{7442859} has independent checkpoint intervals for each level which has several concerns such as overlapping checkpoints which makes it challenging to implement in real-world systems~\cite{8425492}.

\subsection{Our contribution}

This paper is the first to present a theoretical framework for determining optimal parameter settings in a multi-level global checkpointing system that uses a unique periodic checkpoint iinterval, applicable to Exascale stream processing. Our approach is stochastic, where at a given checkpoint interval, a level $l\in{1,2,\dotsc,L}$ is selected with some probability $p_l$ for checkpointing, where $\sum_{l=1}^L p_l=1$. We derive the optimal checkpoint interval, $T^*$, and associated optimal checkpoint probabilities, $p_1^*,p_2^*,\dotsc,p_l^*$ for a multi-level system, that considers as input: failure rates, $\lambda_1 > \lambda_2 > \cdots > \lambda_L$, checkpoint costs $c_1 < c_2 < \cdots < c_L$ and restart costs, $r_1 < r_2 < \cdots < r_L$. Our derivation takes into account higher order effects such as failure during recovery and multiple failures during a single interval. Existing multi-level checkpointing models, such as proposed by Di et al.~\cite{6877346} for HPC applications, minimize the total runtime of an application, given a bounded workload. In stream processing systems this is not applicable. To overcome this, we maximize \emph{utilization}, $0<U<1$, of the system - the fraction of total time available to do useful work, which is applicable to an unbounded workload. We confirm our results with stochastic simulation and practical experimentation.

This paper provides the fundamentals of multi-level checkpointing in section~\ref{sec:probabilistic}, derivation of 2-level checkpointing utilization in Section~\ref{sec:2-lModel}, applies this to a stream processing system in Section~\ref{sec:utilizationStream}, derivation of $L$-level checkpointing utilization in Section~\ref{sec:l-lModel}, applies this to a stream processing system in Section~\ref{sec:LutilizationStream}, shows experimental results in Section~\ref{sec:modelEvaluation}, gives some additional related work in Section~\ref{sec:relatedWork} and some concluding remarks in Section~\ref{sec:conclusion}.

\section{Probabilistic Checkpointing}
\label{sec:probabilistic}
In a single-level checkpointing process, the checkpointing cost and the restart cost remain the same and all failure types are recoverable through the stored checkpoints. For example, failures causing a process to die and failures causing an entire rack to go down are recovered from the same type of checkpoint. Checkpoints are usually stored in an external storage system as checkpoints have to be accessible after any type of a failure.  This results in unnecessary recovery cost for less severe failures which could recover quickly using checkpoints persisted in-memory or machine local storage.

However, multi-level checkpointing consists of different types of checkpoints with varying checkpoint and restart costs. Different types of checkpoints are performed considering different types of failures and their failure rates. In this approach, as the checkpoint level goes up, the failure rate of the level becomes lower and the checkpoint and restart become more expensive. Exiting work focuses on pattern-based multi-level checkpointing, where a pattern is repeated periodically and not necessarily every checkpoint being performed periodically.

In our work, we propose a probabilistic approach of multi-level checkpointing which can easily be adopted by existing stream processing systems. In the probabilistic approach, the system has one global checkpoint interval, while the checkpoint level that should be performed at the checkpoint time is defined by a discrete probability distribution over the levels. We formulate an analytical expression to determine the optimal checkpoint interval and the optimal probability distribution.

\begin{figure}[t]
\centering
\includegraphics[width=3.5in]{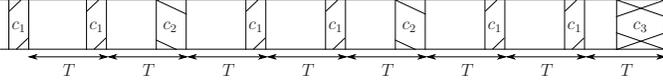}
\caption{Multi-level checkpointing with 3 levels ($c_1 < c_2 < _3$). The cost of the checkpoint operation is shown as a hatched area, the remaining time is useful time availalbe for utilization.}
\label{fig:ml}
\end{figure}

For a multi-level checkpointing process with $L$ levels, let $p_l$ be the probability of performing a level-$l\in\{1,2,\dotsc,L\}$ checkpoint, $c_l$ be the checkpointing cost and $r_l$ be the restart cost of level-$l$. Fig.~\ref{fig:ml} shows a multi-level checkpointing process with a checkpointing periodicity of $T$ using three types of checkpoints with checkpointing probabilities, $p_1=\frac{2}{3}, p_2=\frac{2}{9}, p_3=\frac{1}{9}$ and level-1 with $c_1$ checkpointing cost, level-2 with $c_2$ ($c_1 < c_2$) and level-3 with $c_3$ ($c_2 < c_3$). In this model, failures of any level can be recovered from checkpoints of the same level or any of the higher levels. For a level-3 checkpoint process, level-1 failures can be recovered from any checkpoint, level-2 failures can be recovered from a level-2 or a level-3 checkpoint while level-3 failures can only be recovered from a level-3 checkpoint. As lower level failures can be recovered from the same or higher level checkpoints, if a level-$l$ failure occurs, we assume that the system restores from the nearest completed checkpoint which is of level $l$ or higher. For instance, assume a case where the system performs a level-3 checkpoint and a level-1 failure occurring after the checkpoint. Since the last completed level-3 checkpoint can recover from level-1 failures, the system uses that checkpoint to recover and does not go back to the last completed level-1 checkpoint.

We model utilization of a system to determine the optimal checkpointing interval and optimal probabilities. Utilization, $0 < U < 1$, of a system is defined as the fraction of the system’s time for which its resources are available to do useful work (to process load), as opposed to work done solely to maintain the system’s operation, sometimes called overhead, which in our definition includes the overhead associated with loss and recovery from system failure. The work done by the system to create a checkpoint, and the work done by the system from the checkpoint time to the time taken to detect an occurred failure and successfully restart from the checkpoint is not useful work under our definition and thereby detracts from the utilization. The only useful work is therefore the work done, without failure, between two consecutive checkpoints (not including the work to create the checkpoints), or between a successful restart to the next checkpoint. In the next section, we start with the simplest form of multi-level checkpointing, 2-level checkpointing and subsequently consider $L$-level checkpointing in later sections.

\section{2-level checkpointing Utilization Model}
\label{sec:2-lModel}
In this section, we provide a derivation of our utilization model for 2-level checkpointing from first principles. A 2-level checkpointing process consists of two levels with each level having an independent failure rate, checkpoint cost and restart cost. Level-1 failures can recover from a level-1 or a level-2 checkpoint while level-2 failures can recover only from a level-2 checkpoint. Therefore, we assume that in case of a level-1 failure the system restarts from the last completed checkpoint which could be a level-1 or a level-2 and in case of a level-2 failure, the system restarts from the last completed  level-2 checkpoint. For example, a system can maintain level-1 checkpoints as static in-memory objects which can be used to restore after a thread failure and use an external storage to persist level-2 checkpoints which can recover from more severe failures. In this approach checkpoints stored in both external storage and in-memory can be used to recover from a thread failure and if a machine fails then the state can only be restored from the external storage.

\begin{figure}[t]
\centering
\includegraphics[width=2.5in]{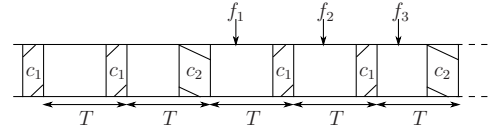}
\caption{2-level checkpointing with period $T$, $p_1=\frac{2}{3}$, overhead $c_1$, $p_2=\frac{1}{3}$, overhead $c_2$ and no failures occurring.}
\label{fig:2l}
\end{figure}

\subsection{Without failure}
Consider a system that performs two-level checkpointing with a constant periodicity of $T$ seconds where the probability of performing a level-1 checkpoint is $p_1$ and the probability of level-2 checkpoint is $p_2$. Let $0 < c_1 < c_2 < T$ be abstract constant costs for level-1 and level-2 checkpointing, here expressed without loss in generality as to how or when the checkpoint is created, in units of time, which without failure as shown in Fig.~\ref{fig:2l} leads to an expression for expected utilization:
\begin{equation}
\label{eq:usimple}
U=\frac{T-p_1c_1-p_2c_2}{T}.
\end{equation}
In our work we depict the checkpoint as being created in the last $c_1$ or $c_2$ seconds of the period $T$ based on the checkpoint level.

\subsection{With failure and negligible restart cost}
\label{sec:failures}
Similar to prior studies~\cite{8425492,7795220,7442859,6877346,Daly:2006:HOE:1134241.1708449}, we model failure of level $l\in\{1,2\}$ as a series of independent failure events having an exponential inter-arrival time, given by failure rate $\lambda_l$,  where the probability of a level-$l$ failure at time $t$ is $\lambda_l e^{-\lambda_l t}$, the probability of failure by time $t$ is $1-e^{-\lambda_l t}$. 
Let $\Lambda=\lambda_1 + \lambda_2$ be the combined failure rate of both failure levels, in which case the probability of failure considering failures of both levels at time $t$ is $\mathbb{P}[\mathbf{X}=t;\Lambda]=\Lambda e^{-\Lambda t}$, the probability of failure by time $t$ is $\mathbb{P}[\mathbf{X}<t;\Lambda]=1-e^{-\Lambda t}$, and the mean time to failure is $\mathbb{E}[\mathbf{X}]=\int_0^\infty t\,\mathbb{P}[\mathbf{X}=t;\Lambda]\,dt = \frac{1}{\Lambda}$.

\begin{figure}[t]  
   \begin{minipage}[t]{0.5\textwidth}
    	\centering
		\includegraphics[height=.7in]{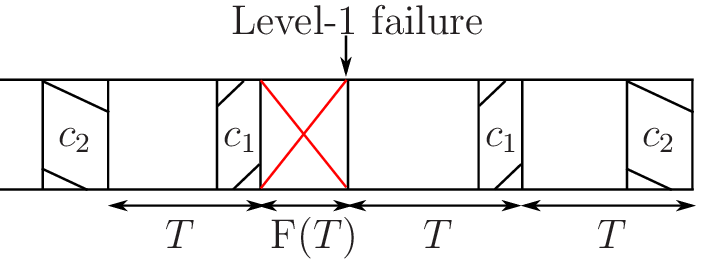}
		\caption{Level-1 failure resulting in lost time $\mathrm{F}(T)$ and instantaneous recovery followed by a successful period $T$.}
		\label{fig:1F}
    \end{minipage}
    \hfill
    \begin{minipage}[t]{0.45\textwidth}  
        \centering 
		\includegraphics[height=.7in]{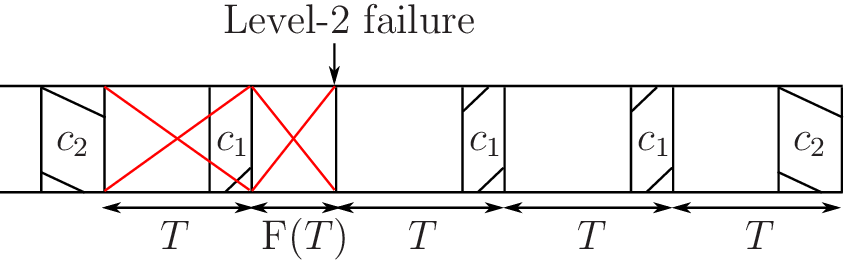}
		\caption{Level-2 failure resulting in lost time $T+\mathrm{F}(T)$ and followed by a successful period $T$.}
		\label{fig:2F}
        \end{minipage}
\end{figure}

Assume that the system has just completed a checkpoint period of either level. Either no failure happens within the next time period $T$, with probability $e^{-\Lambda T}$, in which case the system successfully computes the next checkpoint, or failure of either level happens within time $T$ as indicated in Fig.~\ref{fig:1F},~\ref{fig:2F} and the system needs to restart from an existing checkpoint. Given that a failure of level-1 or level-2 does happen within time $T$, then the mean time to failure, $\mathrm{F}_{\Lambda}(T)$ is:
\begin{equation}
\label{eq:meantime}
\begin{split}
\mathrm{F}_{\Lambda}(T)=\mathbb{E}[\mathbf{X}|\mathbf{X}<T]=\frac{\int_0^T t\,\mathbb{P}[\mathbf{X}=t;\Lambda]\,dt}{\mathbb{P}[\mathbf{X}<T;\Lambda]}\\
=\frac{{\mathrm{e}}^{T\,\Lambda }-T\,\Lambda-1}{\Lambda \,({\mathrm{e}}^{T\,\Lambda }-1)}.
\end{split}
\end{equation}

The value for $\mathrm{F}_{\Lambda}(T)$ gives the average amount of time lost from the last completed checkpoint if a failure happens, not having restarted yet. In this section, for now, we assume the time to detect the failure and to restart is negligible. We include this lost time due to a single failure into our utilization model from \eqref{eq:usimple} by expressing the effective period, $T_{eff}=T+\mathrm{F}_{\Lambda}(T)$, and writing: 
\[U=\frac{T-p_1c_1-p_2c_2}{T_{eff}}.\]
Although failure of either level results in $\mathrm{F}_{\Lambda}(T)$ lost time, level-2 failures can result in additional lost time as depicted in Fig.~\ref{fig:2F}. As level-2 failures can only recover from level-2 checkpoints, the time lost due to a level-2 failure includes the time from the completion of the last level-2 checkpoint and the occurrence of the level-2 failure. This lost time can include zero or more level-1 checkpoints. For instance in Fig.~\ref{fig:2l}, if a level-2 failure occurs in $f_1, f_2, f_3$ then 0, 1, 2 completed level-1 checkpoints are lost respectively in addition to $\mathrm{F}_{\Lambda}(T).$  No checkpoints would be lost if the last completed checkpoint is a level-2 checkpoint which can happen with probabilitiy $p_2$, the probability of loosing a single level-1 checkpoint is $p_1\,p_2$ and the probability of loosing $i$ consecutive level-1 checkpoints is ${p_1}^i\,p_2$, which results in an average $\frac{p_1}{p_2}=\frac{p_1}{(1-p_1)}$ lost completed checkpoints. Therefore, apart from the $\mathrm{F}_{\Lambda}(T)$ lost due to any failure, level 2 failures result in additional $T_{eff}\frac{p_1}{1-p_1}$ lost time, where $T_{eff}$ indicates the average time taken to complete a lost completed level-1 checkpoint.

Figs.~\ref{fig:1F} and~\ref{fig:2F} depict only a single failure occurrence followed by a successful period. In general, the number of consecutive failures that could occur, $k$, before a successful period, is unbounded and selected from the set $k\in\{0,1,2,\dotsc\}$ at random with a geometric distribution, $(1-q)^kq$, having parameter $q=q_{T,\Lambda} =\mathbb{P}[\mathbf{X}\geq T;\Lambda]=1-\mathbb{P}[\mathbf{X}<T;\Lambda]$, giving an average number of consecutive failures of any of the two levels, $\frac{1-q_{T,\Lambda}}{q_{T,\Lambda}}.$ From the average number of consecutive failures, the proportion of level-1 failures is $\frac{\lambda_1}{\Lambda}$ and the proportion of level-2 failures is $\frac{\lambda_2}{\Lambda}$. We can include this lost time into our utilization model by expressing the effective period:
\begin{equation}
\nonumber
\begin{split}
T_{eff}= T+\frac{1-q_{T,\Lambda}}{q_{T,\Lambda}} \bigg(\mathrm{F}_{\Lambda}(T) + \frac{\lambda_2}{\Lambda}\, \Big(\frac{T_{eff}\,p_1}{1-p_1}\Big) \bigg) \\
=\frac{T+\frac{1-q_{T,\Lambda}}{q_{T,\Lambda}}\mathrm{F}_{\Lambda}(T)}{1-\frac{1-q_{T,\Lambda}}{q_{T,\Lambda}}\bigg( \frac{\lambda_2 p_1}{\Lambda(1-p_1)}\bigg)} \qquad \qquad \qquad \qquad 
\end{split}
\end{equation}

\subsection{Including the time to detect and recover from failure}
\label{sec:recover}
\begin{figure}[t]  
   \begin{minipage}[t]{0.5\textwidth}
    	\centering
		\includegraphics[height=.7in]{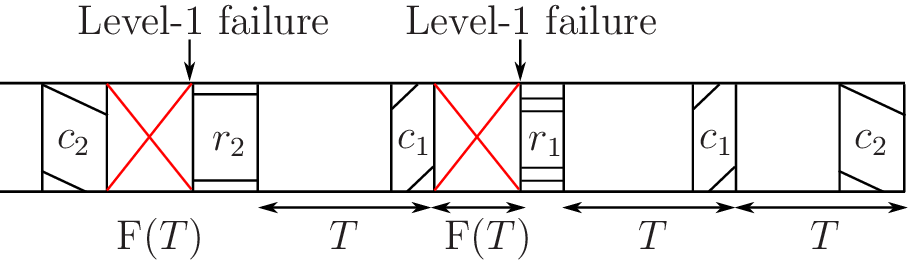}
		\caption{Level-1 failure resulting in lost time $\mathrm{F}(T)$ and recovery time $r_1$ or $r_2$ followed by a successful period $T$.}
		\label{fig:1R}
    \end{minipage}
    \hfill
    \begin{minipage}[t]{0.5\textwidth}  
        \centering 
		\includegraphics[height=.7in]{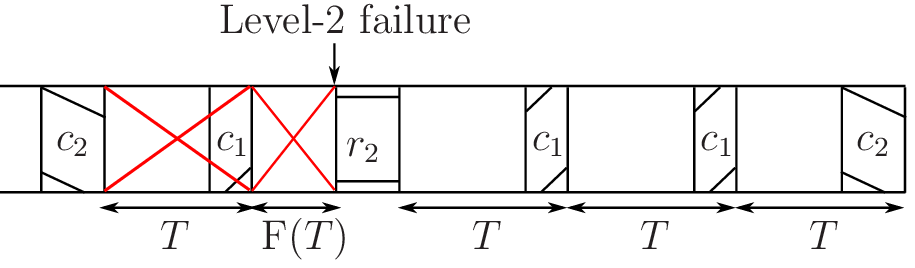}
		\caption{Level-2 failure resulting in lost time $T+\mathrm{F}(T)$ and recovery time $r_2$ followed by a successful period $T$.}
		\label{fig:2R}
        \end{minipage}
\end{figure}

This section includes the time to detect and recover from a failure as shown in Fig.~\ref{fig:1R} and ~\ref{fig:2R}. Every failure requires a restart and the restart cost depends on the failure type. Restart cost $r_1$ is the cost of restarting from a level-1 checkpoint and $r_2$ is the cost of restarting from a level-2 checkpoint. As level-1 failures can recover from either a level-1 or level-2 checkpoint, the restart cost depends on the last completed checkpoint type before the failure. For example, if a level-1 failure occurs after a level-2 checkpoint similar to the first failure shown in Fig.~\ref{fig:1R}, then the restart cost is $r_2$. If a level-1 failure occurs after a level-1 checkpoint as the second failure shown in Fig.~\ref{fig:1R}, then the restart cost is $r_1$. Since the probability of the last completed checkpoint before a level-1 failure being a level-1 is $p_1$, we can write the expected recovery cost of a level-1 failure, $\mathrm{R}_1=p_1r_1 + p_2r_2$. However, for level-2 failures the restart cost is always $r_2$ as level-2 failures can only recover from a level-2 checkpoint which is depicted in Fig.~\ref{fig:2R}. This leads to: 

\begin{equation}
\nonumber
\begin{split}
T_{eff}\!\!=\!T\!+\!\frac{\!1\!\!-\!q_{T,\Lambda}}{q_{T,\Lambda}}\!\bigg(\!\mathrm{F}_{\Lambda}(T) \!+\! \frac{\lambda_1\!(p_1r_1 \!\!+\! p_2r_2)}{\Lambda} \!+\! \frac{\lambda_2}{\Lambda}\!\Big( \frac{T_{eff}p_1}{1\!-\!p_1}\!+\! r_2\Big)\!\!\bigg)\\
=\frac{T+\frac{1-q_{T,\Lambda}}{q_{T,\Lambda}}\bigg(\mathrm{F}_{\Lambda}(T) + \frac{\lambda_1(p_1r_1 + p_2r_2)}{\Lambda} + \frac{\lambda_2r_2}{\Lambda}\bigg)}{1-\frac{1-q_{T,\Lambda}}{q_{T,\Lambda}}\Big( \frac{\lambda_2 p_1}{\Lambda(1-p_1)}\Big)}.\qquad \qquad \quad
\end{split}
\end{equation} 

However, we note that failure may also occur \emph{during} the restart, in which case we assume that the restart must itself start again. Similarly to the number of consecutive failures, the number of attempts to restart, $r\in\{1,2,\dotsc\}$, is selected at random using a geometric distribution (note that following a failure at least 1 restart is always required), $(1-q)^{r-1}q$, with parameter $q$ assuming that the failures of same level, $l$ or lower levels can occur during $r_l$  as restart costs are small and the failure rates of higher levels are very low resulting in almost no higher level failures during $r_l$. Let $q=q_{T,\Lambda_l}=\mathbb{P}[\mathbf{X}\geq r_l;\Lambda_l]$, leading to an average number of restarts $\frac{1}{q_{r_l,\Lambda_l}}\geq 1$, where $\Lambda_l= \sum_{i=1}^{l} \lambda_l.$ For any given failed restart attempt, given that we know the failure occurred within the restart time $r_l$ we know from \eqref{eq:meantime} that the average time lost is $\mathrm{F}_{\Lambda_l}(r_l)$. Therefore, the average number of restarts during $r_1$ is $\frac{1}{q_{r_1,\lambda_1}}$ with an average time lost of $\mathrm{F}_{\lambda_1}(r_1)$, and the average number of restarts during $r_2$ is $\frac{1}{q_{r_1,\lambda_1+\lambda_2}}=\frac{1}{q_{r_1,\Lambda}}$ with an average time loss of $\mathrm{F}_{\lambda_1+\lambda_2}(r_2)=\mathrm{F}_{\Lambda}(r_2).$ Therefore, the recovery cost of a level-1 failure, $\mathrm{R}_1$ and the recovery cost of a level-2 failure, $\mathrm{R}_2$ can be written as:
\[\mathrm{R}_1= p_1\Big(r_1+ \big(\tfrac{1}{q_{r_1,\lambda_1}}-1 \big)\mathrm{F}_{\lambda_1}(r_1)\Big) + p_2\Big(r_2+ \big(\tfrac{1}{q_{r_2,\Lambda}}-1 \big)\mathrm{F}_{\Lambda}(r_2)\Big),\]
\[\mathrm{R}_2=r_2 + \big(\tfrac{1}{q_{r_2,\Lambda}}-1 \big)\mathrm{F}_{\Lambda}(r_2).\]
This leads to:
\begin{equation}
\begin{split}
\nonumber
T_{eff}= T \!+\!\frac{1-q_{T,\Lambda}}{q_{T,\Lambda}} \bigg(\mathrm{F}_{\Lambda}(T) \!+ \frac{\lambda_1}{\Lambda}\,\mathrm{R}_1 \!+ \frac{\lambda_2}{\Lambda}\,\Big( \frac{T_{eff}\,p_1}{1-p_1}+ \mathrm{R}_2\Big)\!\bigg)\\
=\frac{T +\frac{1-q_{T,\Lambda}}{q_{T,\Lambda}} \Big(\mathrm{F}_{\Lambda}(T)\! +\! \frac{\lambda_1\mathrm{R}_1+\lambda_2\mathrm{R}_2}{\Lambda} \Big)}{1-\frac{1-q_{T,\Lambda}}{q_{T,\Lambda}}\Big( \frac{\lambda_2 p_1}{\Lambda(1-p_1)}\Big)},  \quad\quad \quad\quad \qquad \quad \quad  \,
\end{split}
\end{equation}
and finally:
\begin{equation}
\label{eq:simpleSingleU}
\begin{split}
U=\frac{T-p_1c_1-p_2c_2}{T_{eff}} \qquad \qquad  \qquad \qquad \\
=\frac{\Lambda\,\left(\lambda _{2}\,\left(p_{1}\,{\mathrm{e}}^{T\,\Lambda}-1\right)-\lambda _{1}\,p_{2}\right)\,\left(T-c_{1}\,p_{1}-c_{2}\,p_{2}\right)}{\left(1-{\mathrm{e}}^{T\,\Lambda}\right)\,p_{2}\,\left({\mathrm{e}}^{r_{2}\,\Lambda}\,\left(\lambda _{2}+\lambda _{1}\,p_{2}\right)-\lambda _{2}\,p_{1}+\Lambda\,p_{1}\,{\mathrm{e}}^{\lambda _{1}\,r_{1}}\right)}
\end{split}
\end{equation}
This completes the salient features of our 2-level checkpoint and restart system model for a single process. We adapt this to a distributed stream processing system in Section~\ref{sec:utilizationStream}.

\subsection{Optimization of utilization}

Unlike the single level case, that involves just $T$ as the free parameter, deriving closed form equations for the values of $T^*$ and $p_1^*$ that maximize $U$, i.e. by solving $\frac{\partial U}{\partial T}=0$  and $\frac{\partial U}{\partial p_1}=0$, is intractable due to the non-linearities present in the expression for $U$; and more so as we involve more levels.

\begin{figure}[t]
    \centering
        \includegraphics[width=3.1in]{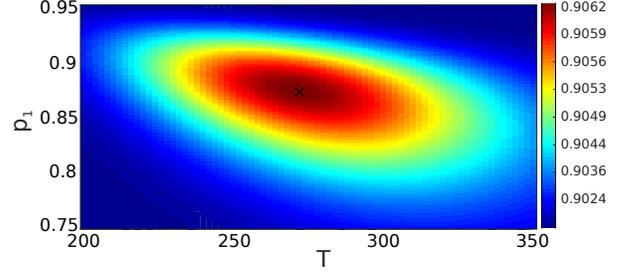}
       \caption{The optimal $U$ sits at the peak of a nearly flat plateau which is shaped by failure rate, checkpoint cost and restart cost ($\lambda_1=24, \lambda_2=0.4$ per day, $c_1=r_1=10$ and $c_2=r_2=30$ seconds).} 
        \label{fig:heatm}
\end{figure}

\begin{figure}[t]
    \centering
        \includegraphics[width=3.6in]{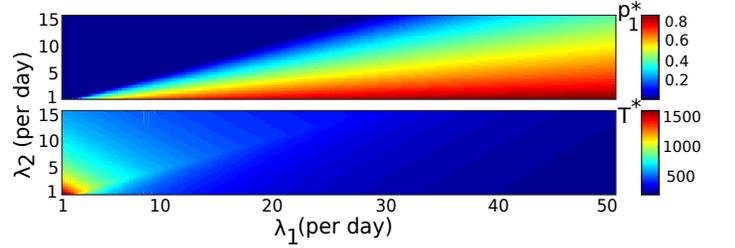}
       \caption{ $T^*$ and ${p_1}^*$ for $c_1\!=\!r_1\!=\!10$ and $c_2\!=\!r_2\!=\!30$ seconds.} 
        \label{fig:optVal}
\end{figure}

Fig.~\ref{fig:heatm} shows how $U$ changes with $T$ and $p_1$ for $\lambda_1=24, \lambda_2=0.4$ per day, $c_1=r_1=10$ and $c_2=r_2=30$ seconds, found using \emph{fmincon} in MATLAB. Utilization is highest in the dark red colored area and the highest utilization is marked by \texttt{x}. 
However, investigating how $T^*$ and $p^*$ vary with varying shape parameters $\lambda_1$ and $\lambda_2$, as shown in Fig.~\ref{fig:optVal}, reveals a regime change when we consider $\lambda_1 >> \lambda_2$. Similar findings appear with respect to shape parameters $r_1$ and $r_2$, i.e. we can consider negligible restart cost. In this regime, which is indeed intuitively the more likely operating regime for a system, we can simplify \eqref{eq:simpleSingleU} and obtain through differentiation:
\begin{equation}
T^*\approx c_{1}\,p_{1}+c_{2}(1-p_{1})+ \frac{{\mathrm{W}}\left(-{\mathrm{e}}^{c_{2}\,\lambda _{1}\,p_{1}-c_{1}\,\lambda _{1}\,p_{1}-c_{2}\,\lambda _{1}-1}\right)+1}{\lambda _{1}}
\label{eq:approxT}
\end{equation}
\begin{equation}
{p_1}^* \approx 1-\sqrt{\frac{\lambda _{2}\,\left(T-c_{1}\right)\,\left({\mathrm{e}}^{T\,\Lambda}-1\right)}{\left(c_{2}-c_{1}\right)\,\left(\lambda _{1}+\lambda _{2}\,{\mathrm{e}}^{T\,\Lambda}\right)}}
\label{eq:approxP}
\end{equation}
where $\mathrm{W}(z)$ is the Lambert $W$ function on the principal branch. These approximations appear to be quite robust, where for all the values of $\lambda_1,\lambda_2$ values shown in Fig.~\ref{fig:optVal}, the maximum difference of the utilization using the actual optimal values and the approximate values given by \eqref{eq:approxT} and \eqref{eq:approxP} is 0.0046.

\subsection{Comparison to 1-level checkpointing}

\begin{figure}[t]
    \centering
    \begin{subfigure}[b]{0.24\textwidth}
        \centering
        \includegraphics[height=1.3in]{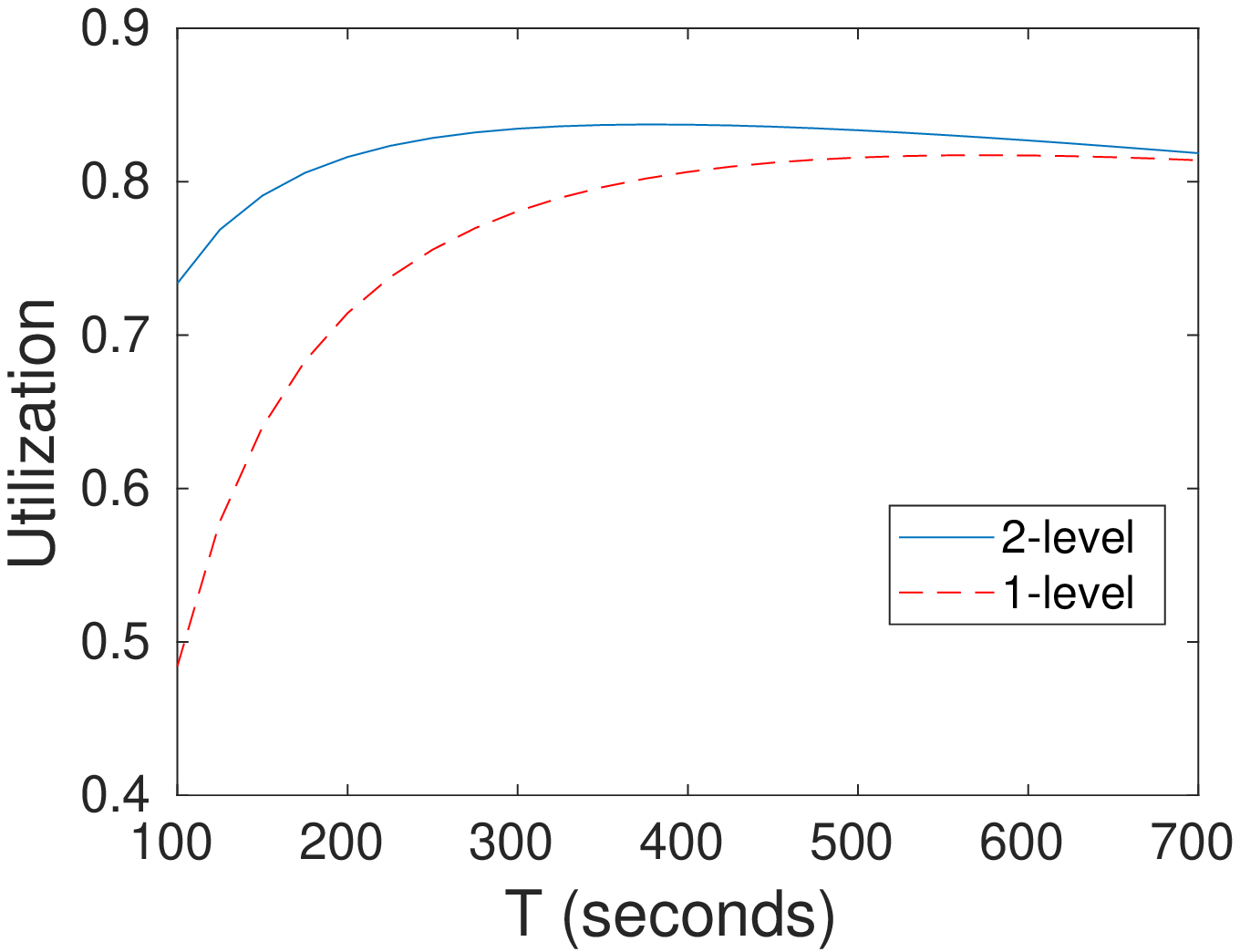}
        \caption{$\lambda_1\!\!=\!\!24,$ $\lambda_2\!\!=\!\!4$ per day, $c_1\!\!=\!\!r_1\!\!=\!\!20,$ $c_2\!\!=\!\!r_2\!\!=\!\!50$ seconds.}    
        \label{fig:1v2.1}
    \end{subfigure}
    \hfill
    \begin{subfigure}[b]{0.24\textwidth}   
        \centering 
        \includegraphics[height=1.3in]{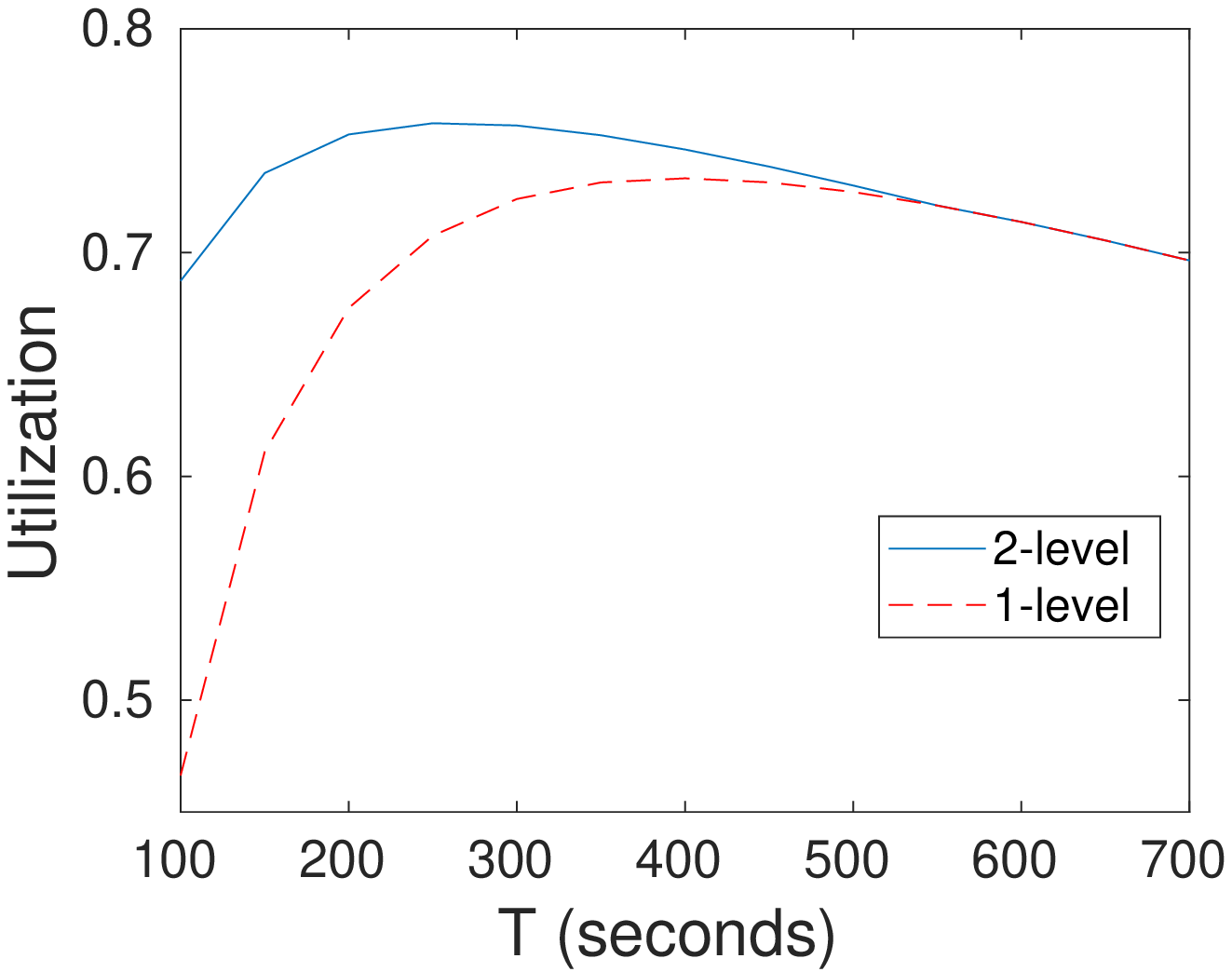}
        \caption{$\lambda_1\!\!=\!\!100,\lambda_2\!\!=\!\!20$ per day, $c_1\!\!=\!\!r_1\!\!=\!\!20,c_2\!\!=\!\!r_2\!\!=\!\!100$ seconds.}     
        \label{fig:1v2.3}
    \end{subfigure}
    \caption{Comparison of 2-level and 1-level checkpointing.} 
    \label{fig:exp1vs2}
\end{figure}

Fig.~\ref{fig:exp1vs2} shows the maximum utilization that can be achieved by a 2-level checkpointing process for different $T$ values, indicated in blue solid lines and the utilization of a 1-level checkpointing process indicated in red dashed lines. In the 1-level process, $p_1=0$, failure rate is $\lambda_1+\lambda_2$ and the system only performs level-2 checkpoints. As shown, using 1-level checkpointing for a system with the same failure rate ($\lambda_1+\lambda_2$) and performing only a single checkpoint type instead of performing 2 levels of checkpoints leads to lower utilization. Furthermore, if $\lambda_2$ is low and $c_2,r_2$ values are comparatively higher than the costs of level-1, then having 2 levels gives noticeable improvements in the optimal utilization as shown in Fig.~\ref{fig:1v2.3}.
Moreover, optimal $p_1$ decreases with $T$ and for large $T$ values optimal $p_1$ comes closer to 0. Therefore, $U$ of 2-level becomes closer to $U$ of 1-level as shown in the figure. 

Table~\ref{tab:1vs2wrtlambda2} shows ${p_1}^*, T^*, U$ for optimal values, maximum $U$ for 1-level where $p_1=0$ and the percentage increase in utilization $\%U$ when using 2-level over 1-level for $\lambda_1=50$ per day, $c_1=r_1=20,$ $c_2=r_2=50$ seconds and different $\lambda_2$ values. As indicated, $p_1$ decreases as $\lambda_2$ increases, and the achievable utilization increase using level-2 checkpointing decreases as the values of $\lambda_2$ and  $\lambda_1$ become closer.

\begin{table}[t]
\centering
\caption{Impact of $\lambda_2$ to ${p_1}^*, T^*,$ and $U.$ }
\label{tab:1vs2wrtlambda2}
\begin{tabular}{|c|c|c|c|c|c|}
\hline
\multirow{2}{*}{$\lambda_2$} & \multirow{2}{*}{${p_1}^*$} & \multirow{2}{*}{$T^*$} & \multirow{2}{*}{$U$}  &  \multirow{2}{*}{\shortstack{$U$ \\ single-level}}&  \multirow{2}{*}{\shortstack{$\%U$ \\ increase}} \\
& & & & &  \\
\hline
0.5 & 0.8897 & 268.0672 & 0.8206 & 0.7549 & 8.6943\\
0.75 & 0.8649 & 268.1357 & 0.8151 & 0.7543 & 8.06\\
1 & 0.8439 & 268.3256 & 0.8106 & 0.7537 & 7.5449\\
5 & 0.6408 &  276.0128 & 0.7712 & 0.7444 & 3.6088\\
10 & 0.4661 & 290.6464 & 0.7448 & 0.7332 & 1.5797\\
\hline
\end{tabular}
\end{table}

\section{2-level Checkpointing for Stream Processing}
\label{sec:utilizationStream}

In this section, we extend the model to work with a distributed stream processing system. Stream processing applications are usually represented as a directed acyclic graph (DAG) of operators where each operator performs computations on the incoming data streams and outputs the results as an output stream to subsequent operators in the DAG. An operator can have multiple operator instances which perform computations on the input data in parallel. Existing stream processing systems use a token based approach to perform checkpointing where a token is sent periodically from source operators till it reaches sink operators, At the arrival of the token, each operator starts performing the checkpoint and passes the token to the next operator or operators. The checkpoint is fully completed once all the operators in the DAG complete the checkpoint. In case of a failure, all the operators restore the last fully completed checkpoint and if there is a partially completed checkpoint, i.e. a checkpoint completed by some of the operators in the DAG is discarded. Furthermore, failure recovery process is the same for any type of failure which could occur in a single operator instance, all the instances of the same operator or set of operators. In our model for stream processing systems, we assume that all operators are stateful and each level of all the operators have same checkpointing costs $0< c_1 < c_2 < T$ and same restart costs, $r_1 < r_2$.

\begin{figure}[t]
        \centering
		\includegraphics[width=3.3in]{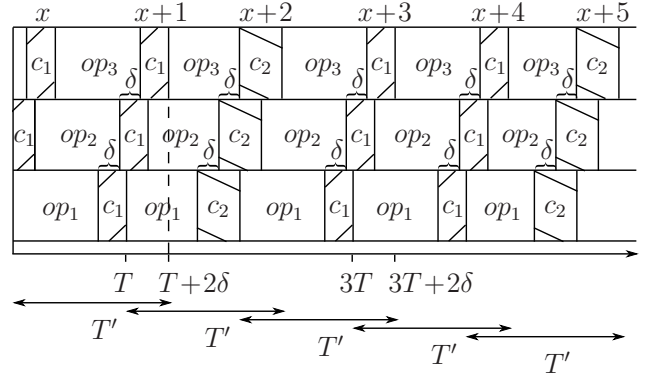}
		\caption{2-level checkpointing of a streaming application with 3 operators. $x$ to $x\!+\!5$ are the ids of completed checkpoints.} 
		\label{fig:strl2}
\end{figure}
Fig.~\ref{fig:strl2} shows 2-level checkpointing for a stream processing application with 3 operators. $\delta$ is the time between sending the token from one operator and receiving the token by the next operator in the DAG. As this DAG consists of 3 operators, the time taken from starting the checkpoint by first operator and completion of the checkpoint by the last operator is $c_1+2\delta$ or $c_2+2\delta.$ Therefore, the time to fully complete a checkpoint from the start of computation is $T+2\delta.$ If the longest path between all source and sink operators (critical path) in the DAG consist of $n$ operators including the source and the sink, then, without considering failure, we have $T_{eff}=T^\prime= T+(n-1)\delta$ and we can write expected utilization as:
\begin{equation} 
U = \frac{T-p_1c_1-p_2c_2}{T_{eff}}=\frac{T-p_1c_1-p_2c_2}{T+(n-1)\delta}.
\label{eq:2}
\end{equation}

\subsection{Including failure and recovery cost}
\label{sec:Strfailures}
In our streaming model, when a level-1 failure occurs then all of the operators restore from the last fully completed checkpoint which could be a level-1 or a level-2 checkpoint. When a level-2 failure occurs then all of the operators restore from the last fully completed level-2 checkpoint. If at least one of the operators in the DAG is still completing a checkpoint at a failure occurrence, that checkpoint is discarded and the system is restored from a fully completed checkpoint.

Similar to sections~\ref{sec:failures} and~\ref{sec:recover}, we include the time due to failures during $T^\prime$ and failures during recovery $(r_1,r_2)$ to improve the accuracy of $T_{eff}$. Similar to the number of consecutive failures, within time $T$ for a single operator the number of attempts to finish $T^\prime$, $k^\prime\in\{1,2,\dotsc\}$, is selected at random using a geometric distribution, $(1-q)^{k^\prime}q$, with parameter $q=q_{T^\prime,\Lambda}=\mathbb{P}[\mathbf{X}\geq T^\prime;\Lambda]=1-\mathbb{P}[\mathbf{X}<T^\prime;\Lambda]$, leading to an average number of consecutive failures $\frac{1-q_{T^\prime,\Lambda}}{p_{T^\prime,\Lambda}}$. Each consecutive failure looses on average an additional $\mathrm{F}_{\Lambda}(T^\prime)$ time. Similarly to the number of consecutive failures, the average number of restarts for recovering from a level-1 checkpoint is $\frac{1}{q_{r_1,\lambda_1}}\geq 1$ and the average time lost due to a failure during restart is $\mathrm{F}_{\lambda_1}(r_1)$. And for recovering from a level-2 checkpoint, the average number of restarts is $\frac{1}{q_{r_2,\lambda_1+\lambda_2}}\geq 1$ and the average time lost due to a failure during restart is $\mathrm{F}_{\lambda_1+\lambda_2}(r_2)$. 
Therefore, the recovery cost of a level-1 failure, $\mathrm{R}_1$ and the recovery cost of a level-2 failure, $\mathrm{R}_2$ can be written as:
\[\mathrm{R}_1 \!=\! p_1\big(r_1+ (\tfrac{1}{q_{r_1,\lambda_1}}-1 )\mathrm{F}_{\lambda_1}(r_1)\big) + p_2\big(r_2+ (\tfrac{1}{q_{r_2,\Lambda}}-1)\mathrm{F}_{\Lambda}(r_2)\big),\]
\[\mathrm{R}_2=r_2 + \big(\tfrac{1}{q_{r_2,\Lambda}}-1 \big)\mathrm{F}_{\Lambda}(r_2).\]
This leads to:
\begin{equation}
\label{eq:streamteff1}
\begin{split}
T_{eff}\!=\!T^\prime+\frac{1-q_{T^\prime,\Lambda}}{q_{T^\prime,\Lambda}} \Big(\mathrm{F}_{\Lambda}(T^\prime) \!+ \frac{\lambda_1}{\Lambda}\,\mathrm{R}_1 \!+ \frac{\lambda_2}{\Lambda}\,\big( \frac{T_{eff} p_1 }{1-p_1}+ \mathrm{R}_2\big)\!\Big) \\
=\frac{T^\prime +\frac{1-q_{T^\prime,\Lambda}}{q_{T^\prime,\Lambda}} \Big(\mathrm{F}_{\Lambda}(T^\prime)\! +\! \frac{\lambda_1\mathrm{R}_1+\lambda_2\mathrm{R}_2}{\Lambda}  \Big)}{1-\frac{1-q_{T^\prime,\Lambda}}{q_{T^\prime,\Lambda}} \Big( \frac{\lambda_2 p_1}{\Lambda(1-p_1)} \Big)}.  \quad\quad \quad\quad \quad \quad\quad \,\,\,
\end{split}
\end{equation}

\subsection{Including the overlap between consecutive checkpoints}
\label{sec:overlap}
Even though the first operator of a DAG completes its checkpoint at $T$, it takes $T^\prime=T+(n-1)\delta$ time for the whole DAG to complete a checkpoint. Therefore, even if the first operator of the DAG starts computation after it finishes its checkpoint, if a failure occurs between the start of the computation and $(n-1)\delta$, then the DAG has to restart from the checkpoint completed by all the operators, not the checkpoint completed only by the first operator. For instance in Fig.~\ref{fig:strl2}, if a level-1 failure occurs during $T$ and $T+2\delta$  then the DAG has to restore from checkpoint $x$ not from $x+1$ as $op_2, op_3$ are not finished with checkpoint $x+1$.  However, if a level-1 failure happens after $T+2\delta$, then the DAG can restore its state from $x+1$ checkpoint.

During the first $(n-1)\delta$ period of $T^\prime$ leading to checkpoint $x+1$, some operators of the DAG are still performing checkpoint $x.$ Therefore, recovery due to failure during the first $(n-1)\delta$ of a $T^\prime$ is the same as if the failure occurs during the previous $T^\prime$. As there is an overlap between the first $(n-1)\delta$ of $T^\prime$ with the previous $T^\prime$ as shown in Fig.~\ref{fig:strl2}, for one $T^\prime$, we only have to consider the time between $(n-1)\delta$ and $T^\prime$, ignoring the first $(n-1)\delta$ of $T^\prime$ which is represented by the previous $T^\prime$. 
The average number of consecutive failures before completing the first $(n-1)\delta$ of $T^\prime$ is $\frac{1-q_{(n-1)\delta, \Lambda}}{q_{(n-1)\delta, \Lambda}}$, where $q=q_{(n-1)\delta, \Lambda}=\mathbb{P}[\mathbf{X}\geq (n-1)\delta;\Lambda]$. Taking all of this into account, the effective period to complete $(n-1)\delta$ is: 
\begin{equation}
\begin{split}
\nonumber
(n-1)\delta + \qquad \qquad \qquad  \qquad \qquad \qquad \qquad \qquad \qquad \qquad \qquad  \\ 
\frac{1-q_{(n-1)\delta, \Lambda}}{q_{(n-1)\delta, \Lambda}}\Big(\mathrm{F}_{\Lambda}\big((n-1)\delta\big)+ \frac{\lambda_1}{\Lambda}\,\mathrm{R}_1 + \frac{\lambda_2}{\Lambda}\,\big( \frac{T_{eff}p_1}{1-p_1}+ \mathrm{R}_2\big)\Big).
\end{split}
\end{equation}
Subtracting this from $T_{eff}$ in \eqref{eq:streamteff1} to avoid the duplicate representation of the overlapping time between two consecutive $T^\prime$ leads to:
\begin{equation}
\begin{split}
U = \frac{T-p_1c_1-p_2c_2}{T_{eff}}={\mathrm{e}}^{-\delta \,\Lambda\,\left(n-1\right)}\,\Lambda\,\left(T-c_{1}\,p_{1}-c_{2}\,p_2\right)\\
\Bigg(\frac{\,\Lambda-\lambda _{1}\,p_{1}-\lambda _{2}\,p_{1}\,\left({\mathrm{e}}^{\Lambda\,\left(T+\delta \,\left(n-1\right)\right)}-{\mathrm{e}}^{\delta \,\Lambda\,\left(n-1\right)}+1\right)}{p_2\,\left({\mathrm{e}}^{T\,\Lambda}-1\right)\,\left({\mathrm{e}}^{r_{2}\,\Lambda}\,\left(\lambda _{2}+\lambda _{1}\,p_2\right)-\lambda _{2}\,p_{1}+\Lambda\,p_{1}\,{\mathrm{e}}^{\lambda _{1}\,r_{1}}\right)}\Bigg).
\label{eq:DAGU}
\end{split}
\end{equation}
This completes the salient features of 2-level checkpointing model for a distributed stream processing system.

\subsection{Optimization of utilization}
The values of $T$ and $p_1$ that maximize the utilization $U$ from \eqref{eq:DAGU}, $T^*$ and ${p_1}^*$ can be found by using nonlinear optimization functions such as \emph{fmincon} in MATLAB. 
Similar to the single process, investigating how $T^*$ and $p^*$ vary with varying shape parameters $\lambda_1$ and $\lambda_2$ reveals a regime change when we consider $\lambda_1 >> \lambda_2$. Similar findings appear with respect to shape parameters $r_1$ and $r_2$, i.e. we can consider negligible restart cost. In this regime, we can simplify \eqref{eq:DAGU} and obtain through differentiation:
\begin{equation}
T^*\approx c_{1}\,p_{1}+c_{2}(1-p_{1})+ \frac{{\mathrm{W}}\left(-{\mathrm{e}}^{c_{2}\,\lambda _{1}\,p_{1}-c_{1}\,\lambda _{1}\,p_{1}-c_{2}\,\lambda _{1}-1}\right)+1}{\lambda _{1}}
\label{eq:StrapproxT}
\end{equation}
\begin{equation}
{p_1}^* \approx 1-{\mathrm{e}}^{\frac{\delta \,\mathrm{n}\,\Lambda}{2}}\,
\sqrt{\frac{\lambda _{2}\,\left(T-c_{1}\right)\,\left({\mathrm{e}}^{T\,\Lambda}-1\right)}{\left(c_{2}-c_{1}\right)\!\!\left(\Lambda\,{\mathrm{e}}^{\delta \,\Lambda}-\lambda _{2}\,{\mathrm{e}}^{\delta \,\mathrm{n}\,\Lambda}+\lambda _{2}\,{\mathrm{e}}^{\Lambda\,\left(T+\delta \,\mathrm{n}\right)}\right)}}
\label{eq:StrapproxP}
\end{equation}
where $\mathrm{W}(z)$ is the Lambert $W$ function on the principal branch. Interestingly approximate $T^*$ is identical to that of a single process, but the approximate ${p_1}^*$ is dependent of $n$ and $\delta.$

We can also consider $T^*$ and ${p_1}^*$ values of a single operator as approximations for $T^*$ and ${p_1}^*$ of a streaming application with same parameters as $\delta$ and $n$ values have a very low influence on $T^*$ and ${p_1}^*.$ For example $T^*$ and ${p_1}^*$ for a single operator with parameters $\lambda_1\!\!=\!\!24, \lambda_2\!\!=\!\!0.4$ per day, $c_1\!\!=\!\!r_1\!\!=\!\!10, c_2\!\!=\!\!r_2\!\!=\!\!30$ seconds is $[271.6709, 0.8737]$ and for same parameters with $\delta=0.5$ seconds and $n=5,n=50,n=500,$  $T^*$ and ${p_1}^*$ values are $[271.6892, 0.8737], [271.6934, 0.8733]$ and $[271.9213, 0.8691]$ respectively.

\section{$L$-level checkpointing Utilization Model}\label{sec:l-lModel}
We now conisder the more general $L$-level checkpointing process, that consists of $L$ failure levels with each level having an independent failure rate, checkpoint cost and restart cost. As the level goes up from 1 to $L$, failure rate goes down ($\lambda_1>\lambda_2> \cdots > \lambda_L$) and checkpoint cost and restart cost goes up ($c_1 < c_2 < \cdots < c_L, r_1 < r_2 < \cdots < r_L,$). In this checkpointing approach, level-$l$ failures can be recovered from checkpoints from level-$l$ to level-$L$. The model assumes that level-$l$ failures are recovered from the latest completed level-$l$ or higher level checkpoint. For example, any level-1 failure can be recovered using any of the checkpoints and therefore, uses the last completed checkpoint to recover from failure. Similar to two-level checkpointing, in $L$-level checkpointing, checkpoints are performed with a constant periodicity of $T$, and probability of performing a level-$l$ checkpoint being $p_l$. Hence without failures we can write the expected utilization as:
\[U=\frac{T-\sum_{l=1}^{L} p_lc_l}{T}\]

\subsection{Including failure and recovery cost}
Similar to sections~\ref{sec:failures} and~\ref{sec:recover}, we include the time due to failures during $T$ and failures during recovery to improve the accuracy of $T_{eff}$. The average number of consecutive failures is $\frac{1-q_{T^\prime,\Lambda}}{p_{T^\prime,\Lambda}}$, where $\Lambda=\sum_{l=1}^{L}\lambda_l$. Each consecutive failure looses on average an additional $\mathrm{F}_{\Lambda}(T)$ time. Similarly to the number of consecutive failures, the average number of restarts for recovering from a level-$l$ checkpoint is $\frac{1}{q_{r_l,\sum_{i=1}^{l}\lambda_i}}\geq 1$, and the average time lost due to a failure during $r_l$ is $\mathrm{F}_{\sum_{i=1}^{l}\lambda_i}(r_l)$.
Therefore, the recovery cost to recover from a level-$l$ failure is:
\[\mathrm{R}_l=\sum_{i=l}^{L} \frac{p_i}{\sum_{j=l}^{L}p_j}\Big(r_i + \big(\frac{1}{q_{r_i,\sum_{j=1}^{i} \lambda_j}}-1\big) \mathrm{F}_{\sum_{j=1}^{i} \lambda_j}(r_i)\Big)\]
where $\frac{p_i}{\sum_{j=l}^{L}p_j}$ indicates the probability of last recoverable checkpoint type being a level-$i$ ($i \geq l$) checkpoint to recover from a level-$l$ failure.
This leads to:
\begin{equation}
\nonumber
T_{eff}= T+\frac{1-q_{T,\Lambda}}{q_{T,\Lambda}} \bigg(\mathrm{F}_{\Lambda}(T) + \sum_{l=1}^{L} \Big(\frac{\lambda_l }{\Lambda} \mathrm{R}_l\Big)\bigg),
\end{equation}
where $\frac{\lambda_l}{\Lambda}$ indicates the proportion of level-$l$ failures from total failures.

Although a failure of any level results in $\mathrm{F}_{\Lambda}(T)$ lost time, failures other than level-1 can result in additional lost time as depicted in Fig.~\ref{fig:2F}. As level-$l$ failures can only recover from a level-$l$ or a higher level checkpoint, the time lost due to a level-$l$ failure includes the time from the completion of the nearest level-$l$ or higher checkpoint and the occurrence of the level-$l$ failure. This lost time can include several lower level checkpoints, i.e. checkpoints from level-1 to level-$(l-1)$. Since the probability of loosing $i$ level-$(l\!-\!1)$ or a lower level checkpoints is ${(\sum_{j=1}^{l-1}p_j)}^i\,(1-\sum_{j=1}^{l-1}p_j)$, on average the number of lost completed checkpoints is: \[\frac{\sum_{i=0}^{l-1}p_l}{1-\sum_{i=0}^{l-1}p_l}=\frac{\sum_{i=0}^{l-1}p_l}{\sum_{i=l}^{L}p_l}.\]
This leads to:
\begin{equation}
\begin{split}
\nonumber
T_{eff}=\! T\!+\!\frac{1-q_{T,\Lambda}}{q_{T,\Lambda}} \bigg(\!\mathrm{F}_{\Lambda}(T) \!+\! \sum_{l=1}^{L}\Big (\frac{ \lambda_l}{\Lambda}\big ( T_{eff}\frac{\sum_{i=0}^{l-1}p_l}{\sum_{i=l}^{L}p_l} + \mathrm{R}_l)\!\Big)\!\bigg) \\
=\frac{T +\frac{1-q_{T,\Lambda}}{q_{T,\Lambda}} \Big(\mathrm{F}_{\Lambda}(T)+ \sum_{l=1}^{L}\frac{\lambda_l\mathrm{R}_l}{\Lambda} \Big)}{1-\frac{1-q_{T,\Lambda}}{q_{T,\Lambda}} \bigg(\sum_{l=1}^{L}\frac{\lambda_l}{\Lambda} \big(\frac{\sum_{i=0}^{l-1}p_l}{\sum_{i=l}^{L}p_l}\big)\!\bigg)}.  \quad\quad  \quad\quad  \quad\quad   \quad\,\,\,    
\end{split}
\end{equation}
and finally:
\begin{equation}
U=\frac{T-\sum_{l=1}^{L} p_lc_l}{T_{eff}}.
\label{eq:L-U}
\end{equation}
This completes the salient features of our $L$-level checkpoint and restart system model for a single process.

\section{$L$-level Checkpointing for Stream Processing}\label{sec:LutilizationStream}
In this section, we extend the $L$-level checkpointing model to a distributed stream processing system. As explained in section \ref{sec:utilizationStream}, if $\delta$ is the time between sending the token from one operator and receiving the token at the next operator in the DAG, and the critical path in the DAG consist of $n$ operators including the source and the sink, then without considering failure, we have $T_{eff}=T^\prime= T+(n-1)\delta$ and we can write the expected utilization as:
\[
U = \frac{T-\sum_{l=1}^{L} p_lc_l}{T_{eff}}=\frac{T-\sum_{l=1}^{L} p_lc_l}{T+(n-1)\delta}.
\]

\subsection{Including failure and recovery cost and the overlap between consecutive checkpoints}\label{subsec:LutilizationStream}
As explained in section~\ref{sec:Strfailures}, we include the time due to failures during $T^\prime= T+(n-1)\delta$ and failures during recovery to improve the accuracy of $T_{eff}$. This leads to:
\begin{equation}
\begin{split}
T_{eff}\!=\!T^\prime\!+\!\frac{1-q_{T^\prime,\Lambda}}{q_{T^\prime,\Lambda}}\!\bigg(\!\mathrm{F}_{\Lambda}(T^\prime) \!+\! \sum_{l=1}^{L}\Big(\frac{\lambda_l}{\Lambda} \big(T_{eff}\frac{\sum_{i=0}^{l-1}p_l}{\sum_{i=l}^{L}p_l} + \mathrm{R}_l\big)\!\Big)\!\!\bigg) \\
=\frac{T^\prime +\frac{1-q_{T^\prime,\Lambda}}{q_{T^\prime,\Lambda}} \Big(\mathrm{F}_{\Lambda}(T^\prime)+ \sum_{l=1}^{L}\frac{\lambda_l\mathrm{R}_l}{\Lambda} \Big)}{1-\frac{1-q_{T^\prime,\Lambda}}{q_{T^\prime,\Lambda}} \bigg(\sum_{l=1}^{L}\frac{\lambda_l}{\Lambda} \big(\frac{\sum_{i=0}^{l-1}p_l}{\sum_{i=l}^{L}p_l}\big)\!\bigg)}.  \quad\quad  \quad\quad  \quad\quad 
\end{split}
\label{eq:Lstr}
\end{equation}
where $\frac{\lambda_l}{\Lambda}$ indicates the proportion of level-$l$ failures from total failures and $\frac{\sum_{i=0}^{l-1}p_l}{\sum_{i=l}^{L}p_l}$ indicates the average number of checkpoints lost due to a level-$l$ failure and $\mathrm{R}_l$ indicates the recovery cost of a level-$l$ failure.

Similar to section \ref{sec:overlap} we have to take into account the overlap between the first $(n-1)\delta$ of $T^\prime$ with the previous $T^\prime$ as shown in Fig.~\ref{fig:strl2}. The effective period to complete $(n-1)\delta$ is:
\begin{equation}
\begin{split}
\nonumber
(n-1)\delta + \qquad \qquad \qquad  \qquad \qquad \qquad \qquad \qquad \qquad \qquad \qquad  \\ 
\frac{1-q_{(n-1)\delta,\Lambda}}{q_{(n-1)\delta,\Lambda}} \bigg(\!\mathrm{F}_{\Lambda}\big((n-1)\delta\big) + \sum_{l=1}^{L}\Big(\frac{\lambda_l}{\Lambda} \big( T_{eff}\frac{\sum_{i=0}^{l-1}p_l}{\sum_{i=l}^{L}p_l} + \mathrm{R}_l\big)\!\Big)\!\bigg)
\end{split}
\end{equation}
We can improve $T_{eff}$ by subtracting the effective period to complete $(n-1)\delta$ from $T_{eff}$ in \eqref{eq:Lstr}. This leads to:
\begin{equation}
\nonumber
\begin{split}
T_{eff}= T  +\frac{1-q_{T^\prime,\Lambda}}{q_{T^\prime,\Lambda}}\,\mathrm{F}_{\Lambda}(T^\prime) 
-\frac{1-q_{(n-1)\delta,\Lambda}}{q_{(n-1)\delta,\Lambda}}\,\mathrm{F}_{\Lambda}\big((n-1)\delta\big) \\
+\Big(\frac{1-q_{T^\prime,\Lambda}}{q_{T^\prime,\Lambda}}-\frac{1-q_{(n-1)\delta,\Lambda}}{q_{(n-1)\delta,\Lambda}}\Big)\sum_{l=1}^{L}\Big(\frac{\lambda_l}{\Lambda}(T_{eff}\frac{\sum_{i=0}^{l-1}p_l}{\sum_{i=l}^{L}p_l} + \mathrm{R}_l)\Big) 
\end{split}
\end{equation}

\begin{equation}
\nonumber
= \frac{
      \splitfrac{
        T  +\frac{1-q_{T^\prime,\Lambda}}{q_{T^\prime,\Lambda}}\, \Big( \mathrm{F}_{\Lambda}(T^\prime) +\sum_{l=1}^{L}\frac{\lambda_l\mathrm{R}_l}{\Lambda}\Big)
      }{
        -\frac{1-q_{(n-1)\delta,\Lambda}}{q_{(n-1)\delta,\Lambda}}\,\Big( \mathrm{F}_{\Lambda}\big((n-1)\delta\big)+\sum_{l=1}^{L}\frac{\lambda_l\mathrm{R}_l}{\Lambda}\Big)
      }
    }{
      1 - \Big(\frac{1-q_{T^\prime,\Lambda}}{q_{T^\prime,\Lambda}}-\frac{1-q_{(n-1)\delta,\Lambda}}{q_{(n-1)\delta,\Lambda}}\Big)\sum_{l=1}^{L}\Big(\frac{\lambda_l}{\Lambda}(T_{eff}\frac{\sum_{i=0}^{l-1}p_l}{\sum_{i=l}^{L}p_l} )\Big)
    }
\end{equation}
and finally: 
\begin{equation}
U=\frac{T-\sum_{l=1}^{L} p_lc_l}{T_{eff}}.
\label{eq:L-UStr}
\end{equation}
This completes the salient features of our $L$-level checkpoint and restart system model for a stream processing system.

\subsection{Comparison to stochastic simulation}
\begin{figure}[t]  
    	\centering 
        \includegraphics[width=1.8in]{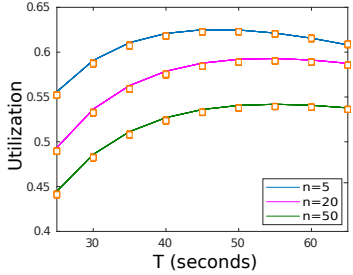}
       \caption{Utilization of 3-level checkpointing for a streaming application for $\lambda_1\!\!=\!\!432,\lambda_2\!\!=\!\!43.2,\lambda_3\!\!=\!\!8.64$ per day, $c_1\!\!=\!\!r_1\!\!=\!\!5,c_2\!\!=\!\!r_2\!\!=\!\!10,c_3\!\!=\!\!r_3\!\!=\!\!20,\delta\!\!=\!\!\!0.5$ seconds and different $n$ values.} 
        \label{fig:3strAll}
    \end{figure}

Fig.~\ref{fig:3strAll} shows the utilization comparison between our model based on \eqref{eq:L-UStr} and the simulation results for 3-level checkpointing using DAGs with different critical path lengths. The solid lines are theoretical utilization while the data points and error bars represent the average utilization and the standard deviation observed after 100 runs, with each simulation running for $\frac{1000}{\lambda_3}$ days. Similarly to 2-level checkpointing, the utilization decreases with the value of $n$ as shown in the figure.

Table \ref{tab:l2vsl3} shows $T^*,{p_1}^*,{p_2}^*$, and $U$ for the optimal values for 3-level checkpointing for $\lambda_1=20,\lambda_2=5$ per day, $c_1=r_1=10,$ $c_2=r_2=20,$ $c_3=r_3=100, \delta=0.5$ seconds and different $\lambda_3$ values along with optimal values and $U$ of 2-level checkpointing for the same parameters and failure rates using $p_2=0$. As indicated in the table, optimal values change slightly as $n$ changes when other parameters are constant. For all the cases level-3 gives better utilization than level-2 checkpointing. The utilization increase we can achieve by 3 levels instead of 2 levels becomes more significant when the difference of $\lambda_3$ and $\lambda_2$ increases or when $n$ increases.

\begin{table}[t]
\centering
\setlength{\tabcolsep}{.4em}
\caption{Comparison of 3-level and 2-level checkpointing}
\label{tab:l2vsl3}
\begin{tabular}{|c|c|c|c|c|c|c|c|c|c|}
\hline
\multirow{2}{*}{$\lambda_3$} & \multirow{2}{*}{$n$} &  \multicolumn{4}{c|}{3-level} & \multicolumn{3}{c|}{2-level} & \multirow{2}{*}{\shortstack{$\%U$ \\ increase}}\\
\cline{3-6}\cline{7-9}
 & & $T^*$ & ${p_1}^*$ & ${p_2}^*$ & $U$ & $T^*$ & ${p_1}^*$ & $U$ & \\
\hline
\multirow{3}{*}{1} & 
  5 & 338.95 & 0.201 & 0.675 & 0.833 & 329.85 & 0.719 & 0.784 & 6.2 \\
& 20 & 339.25 & 0.199 & 0.676 & 0.831 & 330.11 & 0.718 & 0.783 &  6.2\\
& 50 & 339.83 & 0.196 & 0.679 & 0.827 & 330.63 & 0.717 & 0.780 & 6.2\\
\hline
\multirow{3}{*}{0.1} & 
   5 & 336.01 & 0.215 & 0.745 & 0.873 & 322.01 & 0.747 & 0.795 & 9.7\\
& 20 & 336.24 & 0.214 & 0.746 & 0.871 & 322.21 & 0.746 & 0.793 &  9.7\\
& 50 & 336.37 & 0.226 & 0.743 & 0.866 & 322.6 & 0.746 & 0.789 & 9.8\\
\hline
\multirow{3}{*}{0.01} & 
   5 & 335.56 & 0.214 & 0.772 & 0.885 & 321.27 & 0.75 & 0.796 & 11.2\\
& 20 & 335.76 & 0.213 & 0.774 & 0.883 & 321.46 & 0.749 & 0.794 & 11.2\\
& 50 & 336.04 & 0.213 & 0.774 & 0.88 & 321.84 & 0.748 & 0.791 & 11.2\\
\hline
\end{tabular}
\end{table}
\section{Experimental Results with Apache Flink}
\label{sec:modelEvaluation}
We used Apache Flink, a state-of-the-art stream processing framework to evaluate the efficacy of our model. Since Flink does not support multi-level checkpointing, the experiments are for single-level checkpointing. We experimented with various Flink application instances that undertake word counting, a common example used for benchmarking streaming applications, with differing values of topology depth $n$. All experiments were conducted using \emph{m2.medium} nodes running on OpenStack. Each virtual machine had 2 CPU cores, 6 GB of RAM, 30 GB of disk space and ran Ubuntu 15.10, Java 1.7.0\_91, and Flink 1.3.3. We used a five node cluster with one node as master and four nodes as slaves. 
The word count application was loaded with a continuous stream of data and word counting was performed in a sliding window with a set of stateful operators to keep statistics of word counts in each window. Hadoop Distributed File System was used as the state backend for the experiments and Apache Kafka consumer was used as the source. Two methods were used to simulate two levels of random failures. Level-1, throwing an application level exception and level-2, killing a task manager. Throwing an exception and killing a task manager were done based on two exponential distributions at precomputed failure event times with failure rates $\lambda_1,\lambda_2$ respectively. Using different $\lambda_1, \lambda_2$ values we ran each experiment five times, each running for 24 hours. These $\lambda$ values are artificially large so as to indicate results that would be seen at a scale that we cannot experiment with due to lack of resources.

We compared the observed utilization, $\tilde{U}$, obtained using the default Flink parameters for checkpoint interval 30 minutes, with our theoretical prediction of utilization, $U$ using measured parameters $\tilde{c}$, $\tilde{r_1}$ $\tilde{r_2}$ and $\tilde{\delta}$ from Flink's logs as inputs. Although Flink performs single-level checkpoints, we observed that restart cost of the two types of failures we enforced are different. The restart cost from an application level exception ($\tilde{r_1}$) is smaller compared to cost of recovering from a killed task manager ($\tilde{r_2}$) as the later requires more time to restart the task manager. We also computed the  optimal utilization we can achieve if Flink used 2-levels of checkpoints using our theoretical model, taking the cost of level-2 checkpoints as the measured value $\tilde{c}$ from logs and assuming the cost of level-1 checkpoints is half the cost of level-2 checkpoints as measured $\tilde{r_1}$ is approximately half the value of measured $\tilde{r_2}$. Table~\ref{tab:flinkeval} shows the settings of our experiments and observations: $\lambda_1$, $\lambda_2$, $n$, observed $\tilde{c}$, $\tilde{r_1}$, $\tilde{r_2}$ and $\tilde{\delta}$, the observed utilization, $\tilde{U}$, and theoretical utilization, $U$, when $T=30$ minutes, the theoretical optimal, $T^*,{p_1}^*$, for the given settings and observed parameters assuming $c2=\tilde{c}, c1=c2/2$,  theoretical utilization when using the theoretical optimal, $T=T^*, p_1={p_1}^*$, and the percentage increase in utilization $\%U$ over the default $T=30$ minutes if 2-level checkpointing is used. As shown in the table, theoretical predictions of utilization compare well to the observed utilization and we can always achieve significant utilization increase if 2-level checkpointing is used instead of 1-level based on the theoretical model. In the table $T^*$ changes with a change in $n$ because $c$ changes with $n$, as a result of the windowing sizes changing with a deeper topology which increases the checkpoint cost.

\begin{table*}
\centering
\caption{Experimental results using Apache Flink}
\label{tab:flinkeval}
\begin{tabular}{|c|c|c|c|c|c|c|c|c|c|c|c|c|}
\hline
\multirow{2}{*}{$\lambda_1$} & \multirow{2}{*}{$\lambda_2$} & \multirow{2}{*}{$n$} & \multirow{2}{*}{$\tilde{c}$ (ms)} & \multirow{2}{*}{$\tilde{r_1}$ (s)} & \multirow{2}{*}{$\tilde{r_2}$ (s)}  & \multirow{2}{*}{$\tilde{\delta}$ (ms)}   & \multicolumn{2}{c|}{$T=30$ minutes} & \multirow{2}{*}{\shortstack{$T^*$ \\(min)}}  & \multirow{2}{*}{\shortstack{${p_1}^*$ }} & \multicolumn{2}{c|}{$T=T^*\,,\,p_1={p_1}^*$} \\
\cline{8-9}\cline{12-13}
 & & & & & & & $\tilde{U}$ & $U$ & & & $U$ & $\%U$ increase  \\
\hline
\multirow{2}{*}{.02} & \multirow{2}{*}{.008} & 5 & 180.11$\pm$21  & 10.16$\pm$0.01 & 17.51$\pm$1.01 & 12.14$\pm$0.82 & 0.6212$\pm$0.01 & 0.6344 & 0.3805 & 0.3092 & 0.9794 & 57.7\%  \\
& & 7 & 356.21$\pm$20 & 10.18$\pm$0.01 & 17.65$\pm$0.97  & 12.89$\pm$1.04 & 0.6210$\pm$0.01 & 0.6343 & 0.566 & 0.2337 &  0.974 & 56.82\% \\
\hline
\multirow{2}{*}{.03} & \multirow{2}{*}{.008} & 5 & 204.99$\pm$25  & 10.16$\pm$0.01 & 17.56$\pm$0.68 & 11.7$\pm$0.91 & 0.5229$\pm$0.01 & 0.532 & 0.3165 & 0.4715 & 0.9759 & 86.6\%  \\
& & 7 & 427.47$\pm$41 & 10.17$\pm$0.01 & 18.08$\pm$1.11  & 12.97$\pm$0.42 & 0.5186$\pm$0.01 & 0.5319 & 0.4774 & 0.4204 &  0.967 & 86.5\% \\
\hline
\multirow{2}{*}{.04} & \multirow{2}{*}{.01} & 5 & 261.41$\pm$47 & 10.28$\pm$0.17  & 18.04$\pm$1.12 & 13.1$\pm$1.55 & 0.4315$\pm$0.03 & 0.4265 & 0.3062 & 0.4997 & 0.9666 & 124\% \\
&  & 7 & 476.06$\pm$20 & 10.45$\pm$0.39 & 19.47$\pm$3.01 & 13.46$\pm$1.19 & 0.4314$\pm$0.05 & 0.4263 & 0.4228 & 0.4745 & 0.9586 & 122.2\% \\
\hline
\end{tabular}
\end{table*}

\section{Related Work}
\label{sec:relatedWork}

As the checkpoint/restart model is a widely used fault tolerance mechanism, several approaches have been proposed to find the optimal checkpointing frequency which can reduce the impact of the checkpointing on the system performance. For single-level checkpointing, Young~\cite{Young:1974:FOA:361147.361115} has introduced a model to determine the checkpointing frequency that minimizes the time wasted due to failures and Naksinehaboon et al.~\cite{article} also propose approaches focusing on reducing the wasted time. Daly~\cite{Daly:2006:HOE:1134241.1708449} improved Young's model to determine the frequency that gives the minimum total wall clock time to complete an application. Other approaches such as optimal checkpointing frequency approximation based on calculus of variations~\cite{936236,1632007} have also been proposed to reduce the overhead of checkpointing.

Applying the proposed models to stream processing systems is not trivial due to limitations and assumptions made on the models. For example, Jin et al.~\cite{5599253} have proposed a model for HPC environments, which requires the sequential workload of the application. In streaming applications, workload cannot be determined beforehand and such applications can run indefinitely as input streams can be unbounded. The model by Rahman et al.~\cite{7515706} proposed to determine the checkpointing frequency in volunteer computing environments, assumes that a faulty process can only start after a checkpoint interval and the time to detect failure and restart is negligible, which is not the case with streaming applications. Fialho et al.~\cite{5961713} propose a model for uncoordinated checkpointing where each processor performs checkpoints independently, but existing stream processing systems such as Apache Flink and Apache Storm follow a coordinated checkpoint approach.
 
Widely used stream processing systems such as Apache Flink~\cite{Carbone:2017:SMA:3137765.3137777} and Apache Storm use global coordinated periodic checkpoints to support fault tolerance and state management. These systems, facilitate global checkpoints using a token based approach, where a  special token is sent to all the operators in an application from the sources. When a token is received by an operator, it checkpoints the current state and forwards the token to succeeding operators. A global checkpoint is considered as completed, once all the operators complete their checkpoint. Zhuang et al.~\cite{8487327} propose an optimal checkpointing model for stream processing applications by determining an independent checkpointing interval for each operator. However, existing systems use a single checkpointing interval for an application and do not support multiple independent intervals. Furthermore, having independent intervals for each operator requires the system to buffer messages passed between operators to ensure failure recovery. Having buffers for an application with a large number of operators can have a negative impact on application performance.

Most of the existing work done for multi-level checkpointing is for pattern-based checkpoints, where a repeating pattern of checkpoints is defined and in this approach one could define various different patterns for multi-level checkpointing. Moody et al.~\cite{5645453} have presented a Markov model for multi-level checkpointing and shows how multi-level can improve systems efficiency. Benoit et al.~\cite{7795220} have also proposed a similar pattern-based model for HPC systems assuming no failures during checkpoints and recovery, which is not true for real-world systems. Dauwe et al.~\cite{8425492} have derived a model to predict the execution time of applications that uses pattern-based multi-level checkpointing and they determine the optimal settings by trying all the values in their solution space. However, this model is for bounded workloads and therefore, the solution space is bounded which is not the case with streaming applications. 

Di et al.~\cite{6877346} have introduced a mathematical model for multi-level checkpointing in HPC applications which focuses on minimizing the wall clock time. However, this requires the job-length for the optimization which cannot be determined beforehand in a streaming application and assumes failures do not occur during checkpoints and recovery. They have also proposed a 2-level model for HPC applications with unknown lengths assuming that failures do not occur during recovery~\cite{7442859}. They convert the pattern-based solutions to an interval-based solution by providing two independent checkpoint intervals for each level. However, having independent intervals results in overlapping checkpoints and is challenging to implement in real-world systems~\cite{8425492,7795220}.

\section{Conclusion}\label{sec:conclusion}
We presented an analytical expression for the utilization of a distributed stream processing system that uses periodic multi-level checkpoints. This model allows optimization of the checkpoint interval and the probabilities of performing different level checkpoints through maximizing utilization. We have shown the correctness of the model using a stochastic simulation for different levels of checkpoints. With practical experimentation using Flink combined with the theoretical model, we have shown how multi-level checkpointing could improve system efficiency instead of using single-level checkpointing used by well-known stream processing systems.

% or


%\appendices
%% you can choose not to have a title for an appendix
%% if you want by leaving the argument blank
%\section{}

\ifCLASSOPTIONcompsoc
  \section*{Acknowledgments}
\else
  \section*{Acknowledgment}
\fi

This research is funded in part by the Defence Science and Technology Group, Edinburgh, South Australia, under contract MyIP:6104, and was supported by use of the Nectar Research Cloud, a collaborative Australian research platform supported by the National Collaborative Research Infrastructure Strategy (NCRIS).
.

\ifCLASSOPTIONcaptionsoff
  \newpage
\fi


\bibliographystyle{IEEEtran}
\bibliography{reference}
\end{document}